\documentclass[twocolumn,preprintnumbers,showkeys,superscriptaddress,nofootinbib]{revtex4}

\usepackage[dvips]{graphicx,color}
\usepackage{epsfig}
\usepackage{amsmath}
\usepackage{amssymb}
\usepackage{booktabs}
\usepackage[dvipdfm,bookmarks=true,bookmarksnumbered=true,bookmarkstype=toc]{hyperref} 

\allowdisplaybreaks[4]
\newcommand{\lsim}{\raise0.3ex\hbox{$\;<$\kern-0.75em\raise-1.1ex\hbox{$\sim\;$}}}
\newcommand{\gsim}{\raise0.3ex\hbox{$\;>$\kern-0.75em\raise-1.1ex\hbox{$\sim\;$}}}

\newcommand{\uR}{\tilde{u}_R}
\newcommand{\uL}{\tilde{u}_L}
\newcommand{\dR}{\tilde{d}_R}

\newcommand{\nuL}{\tilde{\nu}_L}
\newcommand{\sign}[1]{\mathrm{sign}\big[#1\big]}

\begin{document}

\title{Constraints from Unrealistic Vacua in Supersymmetric Standard
  Model with Neutrino Mass Operators}

\author{Yoshimi Kanehata}
\email{yoshimi@cc.kyoto-su.ac.jp}
\affiliation{Maskawa Institute for Science and Culture, Kyoto Sangyo University, Kyoto 603-8555, Japan}
\affiliation{Department of Physics, Ochanomizu University, Tokyo 112-8610, Japan}

\author{Tatsuo Kobayashi}
\email{kobayash@gauge.scphys.kyoto-u.ac.jp}
\affiliation{Department of Physics, Kyoto University, Kyoto 606-8502, Japan}

\author{Yasufumi Konishi}
\email{konishi@cc.kyoto-su.ac.jp}
\affiliation{Maskawa Institute for Science and Culture, Kyoto Sangyo University, Kyoto 603-8555, Japan}

\author{Takashi Shimomura}
\email{stakashi@yukawa.kyoto-u.ac.jp}
\affiliation{Yukawa Institute for Theoretical Physics, Kyoto University, Kyoto 606-8502, Japan}

\keywords{color and/or charge breaking, unbounded from below, sneutrino, Majorana, seesaw, supersymmetry}
\date{\today}

\preprint{YITP-10-65}
\preprint{KUNS-2285}
\preprint{MISC-2010-07}
\preprint{OCHA-PP-302}

\begin{abstract}
We analyze a scalar potential of the minimal supersymmetric standard
model (MSSM) with neutrino mass operators along
Unbounded-From-Below (UFB) 
and Color and/or Charged Breaking (CCB) directions. We show necessary
conditions to avoid the potential minima which can be 
deeper than the realistic vacuum. 
These conditions would constrain more strongly than conditions in 
the MSSM without taking into account neutrino mass operators, and 
 can improve 
the predictive power of supersymmetric models with neutrino mass operators.
\end{abstract}

\maketitle

\section{Introduction}\label{sec:introduction}
The origin of neutrino masses is one of the unanswered questions in particle physics. It should be addressed by new physics that explains 
tininess of the neutrino masses because the masses of neutrinos are very small compared with other fermions. Small masses are, in general, 
realized by introducing heavy particles. After integrating out the heavy particles, an effective operator which is suppressed by the masses 
of the heavy particles, $M$, is obtained \cite{Weinberg:1979sa},
\begin{align}
  \frac{1}{M}(H \cdot L) (H \cdot L), \label{eq:36}
\end{align}
where $H$ and $L$ are the Higgs and the left-handed lepton doublets. Neutrinos acquire masses through the ElectroWeak Symmetry 
Breaking (EWSB) , and their masses can be very small if $M$ is much larger than vacuum expectation value (vev) of the Higgs. 
The most famous mechanism in this regard is so-called the seesaw mechanism \cite{Minkowski:1977sc,Yanagida:1979,Gell-mann:1979, 
Mohapatra:1979ia,Schechter:1980gr} in which heavy right-handed neutrinos are introduced. One obtains the same mass term, 
Eq.~(\ref{eq:36}), after integrating out the right-handed neutrinos.

Supersymmetric extension of the standard model is one of the promising
candidates for physics above the weak scale.
In supersymmetric extension of the standard model with neutrino mass
operator, Eq.~(\ref{eq:36}), (which we call the $\nu$SSM), there is
the following  operator in the superpotential,
\begin{align}
 c (\hat{H}_2 \cdot \hat{L})(\hat{H}_2 \cdot \hat{L}), \label{eq:37}
\end{align}
and a dimension four operator in the soft supersymmetry (SUSY) breaking term,
\begin{align}
 c' (H_2 \cdot \tilde{L}) (H_2 \cdot \tilde{L}), \label{eq:38}
\end{align}
where $c$ and $c'$ are coupling constants, and $\hat{H_2}~(H_2)$ and $\hat{L}~(\tilde{L})$ are the superfield (the scalar partner) of the 
up-type Higgs and the left-handed leptons. The coupling constants can be determined theoretically once a mechanism for the neutrino 
masses and the SUSY breaking is specified. 
The ratio $c'/c$ would be comparable to (or larger than) 
SUSY breaking masses such as gaugino masses and soft scalar masses.
For example, in the gravity mediation scenario, 
the ratio of the couplings, $|c'|/|c|$, would be on the order of 
 the gravitino mass.

In SUSY models, each of fermions has a scalar partner. The presence of the scalar partners generally leads color and/or charge
breaking (CCB) directions and unbounded-from-below (UFB) directions \cite{Frere:1983ag,AlvarezGaume:1983gj,Derendinger:1983bz,
Kounnas:1983td,Claudson:1983et,Drees:1985ie,Gunion:1987qv,Komatsu:1988mt,Gamberini:1989jw}. 
Along the CCB directions, the scalar potential has minima on which color and/or charge symmetry is spontaneously broken. Along the 
UFB directions, the potential has no global minima and falls down to negative infinity. The existence of these dangerous directions makes 
the vacuum of the EWSB unstable, and hence the these directions must be avoided. In the minimal supersymmetric extension of 
the standard model (MSSM), the scalar potential was analyzed systematically and necessary conditions to avoid the CCB and UFB 
directions were summarized in \cite{Casas:1995pd}. Recently, the scalar potential of the $\nu$SSM with Dirac neutrinos or Majorana 
neutrinos was analyzed along which sneutrinos, scalar partners of the neutrinos, have non-vanishing vev's. It was found in 
\cite{Kobayashi:2010zx} that the UFB directions disappear and turn to CCB directions due to the Yukawa coupling of
neutrinos. Necessary conditions to avoid the CCB directions of the $\nu$SSM were also found in \cite{Kobayashi:2010zx}, which result in 
constraints on the soft SUSY breaking parameters. 
In addition to UFB and CCB directions, false EWSB minima appear 
in the $\nu$SSM.
On such false EWSB minima, either color or charge symmetry 
is not broken, but Higgs scalars and sneutrinos develop their 
large vev's.
Then, such minima would lead to too heavy gauge bosons and 
be excluded by precise electroweak measurements.
Here, we refer such false EWSB minima as CCB minima in view of 
incorrect vacua.

In this article, we consider the $\nu$SSM with the neutrino mass operators, Eq.~(\ref{eq:37}) and (\ref{eq:38}), and derive necessary 
conditions to avoid the CCB and UFB directions. We perform our analysis for tree level potentials, although radiative corrections can modify 
the conditions as discussed in \cite{Ford:1992mv,Casas:1994us,Gioutsos:2002pd}. The conditions obtained in tree-level analysis, however, coincide 
with those from one-loop analysis if the analysis is performed at a scale that the radiative corrections is enough smaller than the tree-level 
potential. We assume that our analysis is performed at this scale.
Also note that our potential is available below $M$, because 
effective operators (\ref{eq:36}), (\ref{eq:37}) and (\ref{eq:38}) 
are induced below $M$.

The outline of this article is organized as follows. In section \ref{sec:constraints-from-ufb}, we briefly review constraints of UFB
and CCB directions in the MSSM. Then we analyze the scalar potential of the $\nu$SSM with the neutrino mass operators and 
derive necessary conditions to avoid UFB and CCB directions in the section \ref{sec:ufb-constr-nussm} and \ref{sec:ccb-constr-nussm}, 
respectively. We show numerical results of constraints on the soft SUSY breaking 
parameters in section \ref{sec:numerical-analysis}. Finally we summarize and discuss our analysis in the section
\ref{sec:summary-discussion}. The scalar potential of the $\nu$SSM with the neutrino mass operators and notations of fields and 
couplings are given in Appendix~\ref{sec:scalar-potential}. In Appendix~\ref{sec:general-form-vacuum}, we give a general form of 
the vacuum expectation values for CCB directions.

\section{Constraints from UFB and CCB directions in the MSSM}\label{sec:constraints-from-ufb}
We start our discussion with briefly reviewing constraints on soft SUSY breaking parameters from UFB and CCB directions in the MSSM. 
According to general properties given in \cite{Casas:1995pd}, there are three types of the UFB and CCB directions respectively, 
which we refer to the "MSSM" UFB and CCB directions in this article. We only show constraints from these directions for comparison 
with our results given in the following sections. Details of the derivation are found in \cite{Casas:1995pd}. Notations of couplings and 
scalar fields are summarized in Appendix \ref{sec:scalar-potential}.

\subsection{The MSSM UFB directions}
The MSSM UFB directions appear along which positive quartic terms in the scalar potential are vanishing or kept under control. 
Along these directions, the potential falls down to negative infinity in large values of fields, making the EWSB vacuum unstable.

The MSSM UFB-1 direction is a direction along which $H_1$ and $H_2$ have an equal non-vanishing vev while other scalars are 
vanishing. The scalar potential along this direction becomes
\begin{align}
 V_{\mathrm{MSSM~UFB-1}} = (m_1^2 + m_2^2 - 2|m_3^2|) |H_2|^2.\label{eq:35}
\end{align}
The potential is unbounded from below unless the coefficient in the right-hand side is positive.
Thus, a necessary condition to be satisfied is 
\begin{align}
 m_1^2 + m_2^2 - 2 |m_3^2| \ge 0. \label{eq:7}
\end{align}
This is the well-known constraint on the soft SUSY breaking masses of Higgses.

Another UFB direction, so-called the MSSM UFB-2 direction, is along which 
\begin{align}
 H_1,~H_2,~\tilde{L} \neq 0, \label{eq:11}
\end{align}
where $\tilde{L}$ is chosen along $\nuL$. The vev's of $H_1$, $H_2$ and $\tilde{L}$ are chosen so that the $D$ term potential is kept 
under control. Then, the potential becomes
\begin{align}
 V_{\mathrm{MSSM~UFB-2}} &= \left( m_2^2 + m_{\tilde{L}}^2 - \frac{|m_3^2|^2}{| m_1^2 - m_{\tilde{L}}^2 |} \right) |H_2|^2 \nonumber \\
                    &\qquad - \frac{2 m_{\tilde{L}}^4}{g_1^2 + g_2^2},
\end{align}
and a necessary condition to avoid a UFB potential is 
\begin{align}
 m_2^2 + m_{\tilde{L}}^2 - \frac{|m_3^2|^2}{| m_1^2 - m_{\tilde{L}}^2 |} \ge 0.
\end{align}

The last direction, the MSSM UFB-3, is along 
\begin{align}
 H_2,~\tilde{L},~\tilde{Q},~\tilde{d}_R \neq 0, \quad \tilde{d}_L = \tilde{d}_R, \label{eq:12}
\end{align}
where $\tilde{Q}$ and $\tilde{L}$ are chosen along $\tilde{d}_L$ and $\nuL$. The vev's of $\tilde{d}_L$ and $\tilde{d}_R$ are chosen 
so that the $F$ term of $H_1$ vanishes. Then, the vev's of $\tilde{d}_L$ and $\tilde{d}_R$ are small compared to those of other scalars 
and can be neglected in the potential. A condition to avoid the MSSM UFB-3 direction is given by 
\begin{align}
 m_2^2 - |\mu|^2 + m_{\tilde{L}}^2 \ge 0. \label{eq:8}
\end{align}
The condition, (\ref{eq:8}), gives a stringent constraint since $m_2^2$ is negative in a large region of parameter space of the MSSM 
for the EWSB to occur.

As stressed in \cite{Kobayashi:2010zx}, the absence of the neutrino Yukawa coupling plays an essential role on 
the MSSM UFB directions, especially the UFB-2 and the UFB-3. It was shown in \cite{Kobayashi:2010zx} that these directions become 
CCB directions when there exits the neutrino Yukawa coupling and the Majorana mass term.

\subsection{The MSSM CCB directions}
The MSSM CCB directions appear along which a negative trilinear term dominates the potential against quartic terms in a certain 
value of fields.

As an example, we consider that $\tilde{Q}$, $\uR$, $\tilde{L}$ as well as $H_1$ and $H_2$ are non-vanishing. 
In order to show constraints from the MSSM CCB directions, it is helpful to express vev's of the scalars in terms of $|H_2|$, 
\begin{align}
 |\tilde{Q}| &= \alpha |H_2|,\quad |\uR| = \beta |H_2|, \nonumber \\
 |H_1| &= \gamma |H_2|, \quad |\tilde{L}| = \gamma_L |H_2|. \label{eq:13}
\end{align}
In the following discussion, we consider that $\tilde{Q}$ is almost a vev along $\uL$ direction. 
Then, the potential can be written 
\begin{align}
 V_{\mathrm{MSSM~CCB}} &= Y_u^2 \alpha^2 \beta^2 \hat{F}(\alpha,\beta,\gamma,\gamma_L) |H_2|^4 \nonumber \\
                       &\quad - 2 Y_u \alpha \beta \hat{A}(\gamma) |H_2|^3 + \hat{m}^2(\alpha,\beta,\gamma,\gamma_L) |H_2|^2, \label{eq:9}
\end{align}
where
\begin{subequations}
\begin{align}
 \hat{F}(\alpha,\beta,\gamma,\gamma_L) &= 1 + \frac{1}{\alpha^2} + \frac{1}{\beta^2} 
                                  + \frac{f(\alpha,\beta,\gamma,\gamma_L)}{\alpha^2 \beta^2}, \\
 f(\alpha,\beta,\gamma,\gamma_L) &= \frac{1}{Y_u^2} \bigg[ 
                                  \frac{1}{8}g_1^2 \left( 1 + \frac{1}{3}\alpha^2 - \frac{4}{3}\beta^2 - \gamma^2 - \gamma_L^2 \right)^2
                                  \nonumber \\
                                 &\quad + \frac{1}{8} g_2^2 (1 - \alpha^2 - \gamma^2 - \gamma_L^2)^2 \nonumber \\
                                 &\quad + \frac{1}{6} g_3^2 (\alpha^2 - \beta^2)^2 \bigg],  \label{eq:13b}\\
 \hat{A}(\gamma) &= |A_u| + |\mu| \gamma, \\
 \hat{m}^2(\alpha,\beta,\gamma,\gamma_L) &= m_1^2 \gamma^2 + m_2^2 -2 |m_3^2| \gamma + m_{\tilde{Q}}^2 \alpha^2   \nonumber \\
                                         &\quad + m_{\uR}^2 \beta^2 + m_{\tilde{L}}^2 \gamma_L^2 .
\end{align}
\end{subequations}
Since the Yukawa couplings of quarks (except for the top) are smaller than the gauge couplings, the deepest direction appears 
along $f(\alpha,\beta,\gamma,\gamma_L)=0$. The extremal value of the up-type Higgs, $|H_2|_{\mathrm{ext}}$, can be obtained 
by solving $\partial V_{\mathrm{MSSM~CCB}} / \partial |H_2| = 0$,
\begin{align}
 |H_2|_{\mathrm{ext}} = \frac{3 \hat{A}}{4 Y_u \alpha \beta \hat{F}} 
    \left( 1 + \sqrt{1 - \frac{8 \hat{m}^2 \hat{F}}{9 \hat{A}^2} } \right). \label{eq:33}
\end{align}
Inserting Eq.~(\ref{eq:33}) into the potential, (\ref{eq:9}), the minimum of the potential is given by
\begin{align}
 V_{\mathrm{MSSM~CCB~min}} &= - \frac{1}{2}\alpha\beta |H_2|_{\mathrm{ext}}^2
   \left( \hat{A} Y_u |H_2|_{\mathrm{ext}} - \frac{\hat{m}^2}{\alpha\beta} \right). \label{eq:34}
\end{align}
The CCB constraints can be obtained by requiring that Eq.~(\ref{eq:34}) is positive.

The MSSM CCB-1 direction is a direction along 
\begin{align}
 &H_2,~\tilde{Q},~\tilde{u}_R \neq 0,\quad |\tilde{d}_L|^2 = |\tilde{d}_R|^2, \label{eq:14}
\end{align}
where $\tilde{d}_L$ and $\tilde{d}_R$ are chosen such that the $F$ term of $H_1$ cancels. Similar to the MSSM UFB-3 
direction, the vev's of $\tilde{d}_L$ and $\tilde{d}_R$ are small and can be neglected in the potential. 
Then, the potential is given by setting $\gamma = 0$, and $\gamma_L^2 = 1-\alpha^2$ with $\alpha = \beta$ for the $D$ term potential 
to vanish. The most stringent constraint to avoid the CCB minimum is given, 
when $m_2^2 - |\mu|^2 + m_{\tilde{L}}^2 > 0$ and $3 m_{\tilde{L}}^2 - (m_{\tilde{Q}}^2 + m_{\uR}^2) + 2 (m_2^2 - |\mu|^2) > 0$,
\begin{align}
 |A_u|^2 \le 3 (m_2^2 - |\mu|^2 + m_{\tilde{Q}}^2 + m_{\uR}^2),
\end{align}
with $\alpha=1$, and when $m_2^2 - |\mu|^2 + m_{\tilde{L}}^2 > 0$ and 
$3 m_{\tilde{L}}^2 - (m_{\tilde{Q}}^2 + m_{\uR}^2) + 2 (m_2^2 - |\mu|^2) < 0$
\begin{align}
 |A_u|^2 &\le \left(1 + \frac{2}{\alpha^2}\right) \nonumber \\
   &\quad \times (m_2^2 - |\mu|^2 + (m_{\tilde{Q}}^2 + m_{\uR}^2)\alpha^2 + m_{\tilde{L}}^2 (1 - \alpha^2)),
\end{align}
with $\alpha^2 = \sqrt{2(m_{\tilde{L}}^2 + m_2^2 - |\mu|^2)/(m_{\tilde{Q}}^2 + m_{\uR}^2 - m_{\tilde{L}}^2)}$.
When $m_2^2 - |\mu|^2 + m_{\tilde{L}}^2 < 0$ and $3 m_{\tilde{L}}^2 - (m_{\tilde{Q}}^2 + m_{\uR}^2) + 2 (m_2^2 - |\mu|^2) < 0$,
the CCB constraint can not be satisfied and the MSSM CCB-1 direction becomes the MSSM UFB-3 direction.

The MSSM CCB-2 direction appears along 
\begin{subequations}
\begin{align}
 &H_1,~H_2,~\tilde{Q},~\uR,~\tilde{L} \neq 0, \\
 &\mathrm{sign}[A_u] = - \mathrm{sign}[B], 
\end{align} \label{eq:15}
\end{subequations}
where $\tilde{Q}$ takes a vev along $\tilde{u}_L$ direction. Similar to the MSSM CCB-1 constraint, the most stringent constraint is obtained as
\begin{align}
 ( |A_u| + |\mu|\gamma)^2 &\le \left( 1 + \frac{2}{\alpha^2} \right) \big( m_2^2 + (m_{\tilde{Q}}^2 + m_{\uR}^2)\alpha^2 \nonumber \\
                            &+ m_1^2 \gamma^2 + m_{\tilde{L}}^2 (1 -\alpha^2 - \gamma^2) - 2 |m_3^2| \gamma \big),\label{eq:10}
\end{align}
with $\alpha = \beta$ and $\gamma_L^2 = 1 - \alpha^2 - \gamma^2$. The minimum of the right-hand side of Eq.~(\ref{eq:10}) can be found 
by numerical calculation by varying $\alpha$ and $\gamma$ between $0$ and $1$.

The MSSM CCB-3 direction is along which the vev's are taken as the same as the MSSM CCB-2 direction but 
$\mathrm{sign}[A_u] = \mathrm{sign}[B]$. The constraint is given by Eq.~(\ref{eq:10}) with the opposite sign of one of the three terms, 
$(|A_u|,~|\mu|\gamma,~-2|m_3^2|\gamma)$.

\section{UFB constraints in the $\nu$SSM with neutrino mass operators}\label{sec:ufb-constr-nussm}
In this and the following sections, we analyze the scalar potential of the $\nu$SSM with neutrino mass operators. As was explained 
in Sec.~\ref{sec:introduction}, the neutrino mass operators consist of 
the $(\hat{H}_2 \cdot \hat{L})(\hat{H}_2 \cdot \hat{L})$ operator 
(\ref{eq:37}) in the superpotential and 
a dimension four operator (\ref{eq:38}) in the soft SUSY breaking term. New terms of $6$th, $5$th and $4$th power of scalars arise from 
these mass operators and modify the structure of the potential. The MSSM UFB directions turn to CCB directions while the MSSM CCB 
directions have another minima under certain conditions that we will discuss later. We focus our analysis on the new minima appearing 
due to these higher power terms and show conditions to avoid them. In the following, we assume that only one sneutrino has non-vanishing vev 
and we take a basis that the couplings of the neutrino mass operators, $c$ and $c'$, are diagonal. The full potential of the model is given in the 
Appendix~\ref{sec:scalar-potential}.

\subsection{Constraints from the MSSM UFB-2 direction}
Let us first consider the MSSM UFB-2 direction which is given by Eq.~(\ref{eq:11}).
The effective scalar potential along the MSSM UFB-2 direction is given 
\begin{align}
  V_{\mathrm{UFB-2}} &= m_1^2 |H_1|^2 + m_2^2 |H_2|^2 - 2 \mathrm{Re}(m_3^2 H_1 H_2) + m_{\tilde{L}}^2 |\tilde{\nu}_L|^2 \nonumber \\
   &\quad + \frac{1}{8}(g_1^2 + g_2^2) ( |\tilde{\nu}_L|^2 + |H_1|^2 - |H_2|^2 )^2 \nonumber \\
   &\quad - 2 \mathrm{Re}\big( c^\ast \mu H_1 H_2^\ast (\tilde{\nu}_L^\ast)^2 \big) 
    - \mathrm{Re}\big( c' (H_2)^2 (\tilde{\nu}_L)^2 \big) \nonumber \\
   &\quad + |c|^2 |\tilde{\nu}_L|^2 |H_2|^2 ( |\tilde{\nu}_L|^2 + |H_2|^2). \label{eq:2}
\end{align}
It is easily understood that the potential is lifted up in large vev's since the term of $6$th power is always 
positive. Thus, the MSSM UFB-2 direction becomes a CCB direction. Note that, along the MSSM UFB-2 direction, neither 
color nor electric charge symmetry is broken, but the Higgses and sneutrinos acquire large vev's on the minima. Such minima 
result in too heavy masses of the gauge bosons and are excluded by precise electroweak measurements. Hence the EWSB does not 
occur correctly on such minima. We refer this false EWSB directions as CCB directions in view of incorrect vacuum.

Before we start detailed analysis, it is helpful to parametrize vev's as, 
\begin{align}
 |\nuL| = \alpha |H_2|,~\quad |H_1| = \gamma |H_2|,
\end{align}
and choose phases of vev's so that terms with undermined phases are negative. This choice of the phases are always possible. 
Then, the potential, (\ref{eq:2}), is expressed as 
\begin{align}
 V_{\mathrm{UFB-2}} &= \hat{C}(\alpha,\gamma) |H_2|^6 - \hat{F}(\alpha,\gamma) |H_2|^4 + \hat{m}^2(\alpha,\gamma) |H_2|^2, \label{eq:3}
\end{align}
where
\begin{subequations}
 \begin{align}
  \hat{C}(\alpha) &= \alpha^2 (\alpha^2 + 1) |c|^2, \\
  \hat{F}(\alpha,\gamma) &= 2 \alpha^2 \gamma |c| |\mu| + \alpha^2 |c'| - f(\alpha,\gamma), \\
  f(\alpha,\gamma) &= \frac{1}{8} (g_1^2 + g_2^2 ) (\alpha^2 + \gamma^2 -1)^2, \label{eq:18} \\
  \hat{m}^2(\alpha,\gamma) &= m_1^2 \gamma ^2 + m_2^2 -2 \gamma |m_3^2| + \alpha^2 m_{\tilde{L}}^2.
 \end{align}
\end{subequations}
Differentiating the potential, (\ref{eq:3}), with respect to $|H_2|$, the extremal value of the up-type Higgs 
is obtained 
\begin{align}
 |H_2|_{\mathrm{ext}}^2 = \frac{\hat{F}(\alpha,\gamma)}{3 \hat{C}(\alpha)} 
  \left( 1 + \sqrt{ 1 - \frac{3 \hat{C}(\alpha) \hat{m}^2(\alpha,\gamma)}{\hat{F}^2(\alpha,\gamma)} } \right),\label{eq:4}
\end{align}
and then the minimum of the potential becomes
\begin{align}
 V_{\mathrm{UFB-2~min}} &= -\frac{1}{3} \hat{F}(\alpha,\gamma) |H_2|_{\mathrm{ext}}^2 
  \left( |H_2|^2_{\mathrm{ext}} - \frac{2 \hat{m}^2(\alpha,\gamma)}{\hat{F}(\alpha,\gamma)} \right). \label{eq:5}
\end{align}
The minimum becomes the deepest when $f(\alpha,\gamma)=0$, imposing $\alpha^2 = 1 - \gamma^2$. The typical order of 
$|H_2|_{\mathrm{ext}}$ is 
\begin{align}
 |H_2|_{\mathrm{ext}} \sim \sqrt{ m_{\mathrm{SUSY}} M },
\end{align}
where $m_{\mathrm{SUSY}}$ and $M$ are a typical scale of the soft SUSY breaking masses and a cut-off scale of the neutrino mass operators. 
Thus, the potential could be deeper than that of the EWSB if $M$ is larger than $m_{\mathrm{SUSY}}$. A necessary condition to avoid 
this CCB minimum requires that the minimum of the potential, (\ref{eq:5}),  becomes positive, i.e. 
\begin{align}
 \left( \frac{|c'|}{|c|} + 2 |\mu| \gamma \right)^2 \le \frac{4 (2 - \gamma^2) }{1 - \gamma^2} \hat{m}^2(\gamma), \label{eq:6}
\end{align}
where $\hat{m}^2(\gamma)$ is the one inserting $\alpha^2 = 1 -
\gamma^2$. 
Then, the most stringent constraint is obtained by minimizing  
the following function $\eta(\gamma)$ with respect to $\gamma$,
\begin{align}
\eta(\gamma) = 2\sqrt{\frac{ 2 - \gamma^2 }{1 - \gamma^2}
  \hat{m}^2(\gamma)} - 2 |\mu| \gamma .
\end{align}
That is, the extremal value, $\gamma_{\mathrm{ext}}$, is given as a solution of 
the function,
\begin{align}
 \xi(\gamma)=&\gamma \hat{m}^2(\gamma) + ( 1 - \gamma^2) (2 - \gamma^2) \big( (m_1^2 - m_{\tilde{L}}^2) \gamma - |m_3^2| \big) \nonumber \\ 
  &\quad - (1-\gamma^2)\sqrt{(1-\gamma^2) (2 - \gamma^2) \hat{m}^2(\gamma)} |\mu|, \label{eq:16}
\end{align}
where $\xi(\gamma) \propto \partial \eta(\gamma)/\partial \gamma$.
One can see that the constraint, (\ref{eq:6}), is similar to the constraint, (\ref{eq:10}).
 
\subsection{Constraints from the MSSM UFB-3 direction}
Next we consider the MSSM UFB-3 direction defined in Eq.~(\ref{eq:12}). Parameterizing the vev's as 
\begin{align}
 |\nuL| = \alpha |H_2|,
\end{align} 
the scalar potential along this direction is given in the same form as Eq.~(\ref{eq:3}) by setting $\gamma=0$ and making 
a replacement $m_2^2$ with $m_2^2 - |\mu|^2$. The minimum of the potential is obtained for $\alpha = 1$ and the most stringent 
constraint is given by using Eqs.~(\ref{eq:4}) and (\ref{eq:5}),
\begin{align}
 \frac{|c'|^2}{|c|^2} \le 8( m_2^2 - |\mu|^2 + m_{\tilde{L}}^2 ). \label{eq:17}
\end{align}
Comparing the constraint, (\ref{eq:17}),  with that in the MSSM, (\ref{eq:8}), we can see that Eq.~(\ref{eq:17}) imposes more sever 
bound on the soft SUSY breaking parameters if the ratio, $|c'|/|c|$, is of order the SUSY breaking scale.

\section{CCB Constraints in the $\nu$SSM with neutrino mass operators}\label{sec:ccb-constr-nussm}
We analyze the potential along the MSSM CCB direction. We focus new minima which occur due to the higher order 
terms of $6$th, $5$th and $4$th power of fields. The scalar potential with the parametrization of Eq.~(\ref{eq:13}) is expressed as 
\begin{align}
 V &= \hat{C}(\gamma_L) |H_2|^6 -  \hat{D}(\alpha,\beta,\gamma_L) |H_2|^5 + \hat{F}(\alpha,\beta,\gamma,\gamma_L) |H_2|^4 \nonumber \\
   &\quad - \hat{A}(\alpha,\beta,\gamma) |H_2|^3 + \hat{m}^2(\alpha,\beta,\gamma,\gamma_L) |H_2|^2,\label{eq:27}
\end{align}
where 
\begin{subequations}
\begin{align}
 \hat{C}(\gamma_L) &= \gamma_L^2 ( 1 + \gamma_L^2 ) |c|^2, \\
 \hat{D}(\alpha,\beta,\gamma_L) &= - 2 \epsilon_1 \alpha \beta \gamma_L^2 |Y_u| |c|, \\
 \hat{F}(\alpha,\beta,\gamma,\gamma_L) &= |Y_u|^2 (\alpha^2 \beta^2 + \alpha^2 + \beta^2) \nonumber \\
                                       & - \gamma_L^2 ( \epsilon_2 |c'| + 2 \epsilon_3 \gamma |\mu| |c| )  \nonumber \\
                                       & + |Y_u|^2 f(\alpha,\beta,\gamma,\gamma_L), \\
 \hat{A}(\alpha,\beta,\gamma) &= 2 \alpha \beta |Y_u| \big( \epsilon_4 |A_u| + \epsilon_5  \gamma |\mu| \big), \\
 \hat{m}^2(\alpha,\beta,\gamma,\gamma_L) &= \gamma^2 m_1^2 + m^2_2 - 2 \epsilon_6 \gamma |m_3^2| \nonumber \\
                                         &\quad + \alpha^2 m^2_{\tilde{Q}_L} + \beta^2 m^2_{\uR} + \gamma_L^2 m^2_{\tilde{L}},
\end{align}
\label{eq:19}
\end{subequations}
and $f(\alpha,\beta,\gamma,\gamma_L)$ is given in Eq.~(\ref{eq:13b}). 
Here, $\epsilon_i$ $(i=1-6)$ denote the sign ($\pm$) and 
are defined as
\begin{subequations}
\begin{align}
 \epsilon_1 &= \sign{\mathrm{Re}\big( Y_u c^\ast \uL \uR^\ast H_2^\ast (\nuL^\ast)^2 \big)}, \\
 \epsilon_2 &= \sign{\mathrm{Re}\big( c' (H_2)^2 (\nuL)^2 \big)}, \\
 \epsilon_3 &= \sign{\mathrm{Re}\big( \mu c^\ast H_1 H_2^\ast (\nuL^\ast)^2 \big)}, \\
 \epsilon_4 &= \sign{\mathrm{Re}\big( A_u Y_u H_2 \uL \uR^\ast \big)}, \\
 \epsilon_5 &= \sign{\mathrm{Re}\big( Y_u \mu^\ast H_1^\ast \tilde{u}_L \tilde{u}_R^\ast \big)}, \\
 \epsilon_6 &= \sign{\mathrm{Re}\big( m_3^2 H_1 H_2 \big)}.
\end{align}
\end{subequations}
In the following, we show possible choices of $\epsilon_i,~(i=1-6)$ for the MSSM CCB-1 and CCB-2 directions and derive 
constraints from the deepest directions.

\subsection{Constraints from the MSSM CCB-1 directions}
The MSSM CCB-1 direction is defined in Eq.~(\ref{eq:14}) and the deepest minimum emerges along which the $D$ term potential is vanishing.
The scalar potential along this direction is obtained from Eqs.~(\ref{eq:19}) 
by setting $\gamma=0$ and replacing $m_2^2$ with $m_2^2-|\mu|^2$,
\begin{subequations}
\begin{align}
 \hat{C}(\alpha) &= (1 - \alpha^2 ) ( 2 - \alpha^2 ) |c|^2, \\
 \hat{D}(\alpha) &= - 2 \epsilon_1 \alpha^2 (1 - \alpha^2) |Y_u| |c|, \\
 \hat{F}(\alpha) &= |Y_u|^2 \alpha^2 (2 + \alpha^2) - \epsilon_2 ( 1 - \alpha^2) |c'|, \\
 \hat{A}(\alpha) &= 2 \epsilon_4 \alpha^2 |Y_u| |A_u|, \\
 \hat{m}^2(\alpha) &= m_2^2-|\mu|^2 + \alpha^2 (m^2_{\tilde{Q}_L} +  m^2_{\uR} ) + ( 1 - \alpha^2 ) m^2_{\tilde{L}},
\end{align}
\end{subequations}
where $\alpha = \beta$ and $\gamma_L^2 = 1 - \alpha^2$ are used. By choosing an appropriate phase of fields, the signs, $\epsilon_{1,2,4}$, 
satisfy a relation 
\begin{align}
 \epsilon_1 \epsilon_2 \epsilon_4 = \mathrm{sign}[c] \mathrm{sign}[c'] \mathrm{sign}[A_u], \label{eq:21}
\end{align}
where we assumed that $c$, $c'$ and $Y_u$, $A_u$ are real numbers. We can find the properties from the relation,
\begin{enumerate}
 \item When the right-hand side of Eq.~(\ref{eq:21}) is positive, all of or one of the three signs can be made positive.
 \item When the right-hand side of Eq.~(\ref{eq:21}) is negative, all of or one of the three signs can be made negative.
\end{enumerate}
The deepest direction corresponds to the choice of signs 
such that $\epsilon_1$ is negative while $\epsilon_2$ and $\epsilon_4$ are positive.

Before we start our analysis, it is important to notice that the terms of $\hat{C}$, $\hat{D}$ and $\hat{F}$ except for $-(1-\alpha^2)|c'|$ 
originate from the $F$ term potential, and therefore the total contribution of the $6$th, $5$th and $4$th order terms is positive if $|c'|/|Y_u^2| \ll 1$. 
In this case, the condition to avoid the MSSM CCB-1 direction is the same as the one in \cite{Casas:1995pd}. The situation, however, 
changes when $|c'|/|Y_u^2| \gg 1$. The $4$th order term is dominated by $-(1-\alpha^2)|c'|$ and the new CCB minima emerge at large values 
of fields. 

As shown in the Appendix~\ref{sec:general-form-vacuum}, the leading terms of $|H_2|_{\mathrm{ext}}$ are independent of the Yukawa 
couplings in the case of $|c'|/|Y_u|^2 \gg 1$. Therefore we can neglect the terms proportional to the Yukawa coupling . Then, the scalar 
potential is approximated as 
\begin{align}
 V_{\mathrm{CCB-1}} \simeq \hat{C}(\alpha) |H_2|^6 + \hat{F}(\alpha) |H_2|^4 + \hat{m}^2(\alpha) |H_2|^2, \label{eq:22}
\end{align}
where
\begin{align}
 \hat{F}(\alpha) = - (1-\alpha^2) |c'|. \label{eq:23}
\end{align}
The extremal value of the up-type Higgs is obtained by differentiating Eq.~(\ref{eq:22}) with respect to $|H_2|$,
\begin{align}
 |H_2|_{\mathrm{ext}}^2 = - \frac{\hat{F}(\alpha)}{3 \hat{C}(\alpha)}\left( 1 - \sqrt{1- \frac{3 \hat{C}(\alpha)
 \hat{m}^2(\alpha)}{\hat{F}^2(\alpha)}} \right), \label{eq:24}
\end{align}
where $\hat{F}(\alpha)$ must be negative for the potential to be minimum. The minimum of the potential is given by 
\begin{align}
 V_{\mathrm{CCB-1~min}} &= \frac{1}{3} |H_2|^2_{\mathrm{ext}} 
  \big( \hat{F}(\alpha) |H_2|^2_{\mathrm{ext}} + 2 \hat{m}^2(\alpha) \big), \label{eq:32}
\end{align}
and a necessary condition to avoid the MSSM CCB-1 minimum is 
\begin{align}
\hat{F}^2(\alpha) < 4 \hat{C}(\alpha) \hat{m}^2(\alpha),\label{eq:25}
\end{align}
which gives the constraint,
\begin{align}
 \frac{|c'|^2}{|c|^2} < 4 \frac{2-\alpha^2}{1-\alpha^2} \hat{m}^2(\alpha) . \label{eq:26}
\end{align}
The most stringent condition is obtained by minimizing the right-hand side of Eq.~(\ref{eq:26}).

\subsection{Constraints from the MSSM CCB-2 directions}
The MSSM CCB-2 direction is defined in Eq.~(\ref{eq:15}). Similar to the MSSM CCB-1 direction, the deepest direction emerges 
along a direction, $\alpha = \beta$ and $\gamma_L^2 = 1 - \alpha^2 - \gamma^2$. Along this direction, the signs, $\epsilon_i,~(i=1-6)$, 
satisfy the relations,
\begin{subequations}
\begin{align}
 &\epsilon_1 \epsilon_2 \epsilon_4 = \mathrm{sign}[c] \mathrm{sign}[c'] \mathrm{sign}[A_u], \\
 &\epsilon_1 \epsilon_3 = \epsilon_5, \\
 &\epsilon_4 \epsilon_5 \epsilon_6 = \mathrm{sign}[A_u] \mathrm{sign}[m_3^2 / \mu].
\end{align} \label{eq:20}
\end{subequations}
The minimum of the potential becomes the deepest when $\epsilon_1$ is negative while the other $\epsilon$'s are positive. 
From Eqs.~(\ref{eq:20}), we can find the following properties,
\begin{enumerate}
 \item $\epsilon_2$ can be always set positive.
 \item When $\mathrm{sign}[A_u] = \mathrm{sign}[m_3^2/\mu]$, $\epsilon_4,~\epsilon_5$ and $\epsilon_6$ can be made positive simultaneously, 
       and $\epsilon_1$ and $\epsilon_3$ must be positive (negative) if $\mathrm{sign}[c] \mathrm{sign}[c'] \mathrm{sign}[A_u]$ is positive (negative).
 \item When $\sign{A_u} = - \sign{m_3^2/\mu}$, two of $\epsilon_4,~\epsilon_5$ and $\epsilon_6$ can be made positive and the other must be negative.
       \begin{enumerate}
	\item  If $\epsilon_4$ and $\epsilon_6$ are positive, $\epsilon_1$ and $\epsilon_3$
	must be the opposite signs each other.
	\item  If either $\epsilon_4$ or $\epsilon_6$ is negative, $\epsilon_1$ and $\epsilon_3$ must be the same sign.
       \end{enumerate}
\end{enumerate}

In the following, we consider the case that $\epsilon_i$ $(i=1-6)$ are all positive. Then, the potential is given by 
\begin{subequations}
\begin{align}
 \hat{C}(\alpha,\gamma) &= ( 1-\alpha^2-\gamma^2 )( 2-\alpha^2-\gamma^2 ) |c|^2, \\
 \hat{D}(\alpha,\gamma) &= - 2 \alpha^2 (1-\alpha^2-\gamma^2) |Y_u| |c|, \\
 \hat{F}(\alpha,\gamma) &= |Y_u|^2 \alpha^2 (\alpha^2 + 2) \nonumber \\
                 &\quad - (1 - \alpha^2 -\gamma^2) (|c'|+2\gamma |\mu| |c|), \\
 \hat{A}(\alpha,\gamma) &= 2 \alpha^2 |Y_u| ( |A_u| + \gamma |\mu| ), \\
 \hat{m}^2(\alpha,\gamma) &= \gamma^2 m^2_{1} + m^2_{2} - 2 \gamma |m^2_3| \nonumber \\
                   &\quad + \alpha^2 (m^2_{\tilde{Q}_L} + m^2_{\uR} ) + (1 - \alpha^2 - \gamma^2) m^2_{\tilde{L}}.
\end{align}
\end{subequations}
Similar to the MSSM CCB-1 direction, a new CCB minimum appears when $|c'|/|Y_u|^2 \gg 1$. The extremal value of $|H_2|$ 
at the leading order is given by Eq.~(\ref{eq:24}) with $\hat{F}(\alpha,\gamma) \simeq - (1 - \alpha^2 -\gamma^2) (|c'|+2\gamma |\mu| |c|)$, 
and a necessary condition to avoid the CCB minimum is 
\begin{align}
  \left( \frac{|c'|}{|c|} + 2 \gamma |\mu| \right)^2 <  4 \frac{2-\alpha^2-\gamma^2}{1-\alpha^2-\gamma^2} \hat{m}^2(\alpha,\gamma). \label{eq:60}
\end{align}

In the end, we comment on other possibilities of choice of vev's. Since the terms proportional to the Yukawa coupling is irrelevant in the present
analysis, similar results can be obtained when we take vev's of squarks as 
\begin{align}
 \tilde{u}_L  \rightarrow \tilde{d}_L, \quad \tilde{u}_R \rightarrow  \tilde{d}_R,
\end{align}
where $\tilde{d}_L$ and $\tilde{d}_R$ are different squarks from those to cancel the $F$ term of $H_1$. 
Along this direction, the constrain for the MSSM CCB-1 is obtained by 
\begin{align}
 \frac{|c'|^2}{|c|^2} < 4 \frac{2+\alpha^2}{1+\alpha^2} \hat{m}^2(\alpha) ,
\end{align}
like Eq.~(\ref{eq:26}), where $\hat{m}^2(\alpha)$ is replaced by 
\begin{align}
\hat{m}^2(\alpha) = m^2_2 - |\mu|^2 + \alpha^2 (m^2_{\tilde{Q}_L} +
m^2_{\dR} ) + ( 1 +  \alpha^2 ) m^2_{\tilde{L}}.
\end{align}
 Similarly, the constraint from the MSSM CCB-2 is obtained by 
\begin{align}
  \left( \frac{|c'|}{|c|} + 2 \gamma |\mu| \right)^2 <  4 \frac{2+\alpha^2-\gamma^2}{1+\alpha^2-\gamma^2} \hat{m}^2(\alpha,\gamma),
\end{align}
like Eq.~(\ref{eq:60}), where $\hat{m}^2(\alpha,\gamma)$ is replaced by 
\begin{align}
 \hat{m}^2(\alpha,\gamma) &= \gamma^2 m^2_{1} + m^2_{2} - 2 \gamma
 |m^2_3| + \alpha^2 (m^2_{\tilde{Q}_L} + m^2_{\dR} )
\nonumber \\
                   &\quad + (1 + \alpha^2 - \gamma^2) m^2_{\tilde{L}}.
\end{align}

\section{Numerical Analysis}\label{sec:numerical-analysis}
In this section, we show numerical results of the constraints of the MSSM UFB-2 and the MSSM UFB-3 derived in the previous sections. 
Similar analysis can be carried out for the MSSM CCB directions.
As an illustrating example, we employ the constrained MSSM (CMSSM) to calculate the soft SUSY breaking parameters and the supersymmetric Higgs masses. 
For the couplings of the neutrino mass operators, we assume that $c$
and $c'$ are so small that these do not contribute to the
renormalization group equations (RGEs) of other couplings and SUSY
breaking parameters, significantly. 
Then, we treat the couplings, $c$ and $c'$,  as input parameters and set values at a scale we perform the numerical calculation. As we mentioned in the 
Sec.\ref{sec:introduction}, the ratio of the couplings of the neutrino mass operators in the minimal supergravity SUSY breaking model is 
expected to be 
\begin{align}
 \frac{|c'|}{|c|} ={\cal O} (m_{3/2}),
\end{align}
where $m_{3/2}$ is the gravitino mass. 
In the following, we 
consider the case that the ratio, $|c'|/|c|$, is between $100$ and $1000$ GeV. We will see that the constraint of the MSSM UFB-3 imposes 
more stringent bound on the soft SUSY parameters than those by the MSSM, Eq.~(\ref{eq:8}) in this case.

The CMSSM is parametrized by four parameters and a sign,
\begin{align}
 {M_{1/2},~m_0,~A_0,~\tan\beta,~\mathrm{sign}[\mu]},
\end{align}
where the first three parameters are the universal gaugino and scalar
masses, and the universal trilinear couplings defined at the grand
unified theory (GUT) scale. 
$\tan\beta$ is the ratio of the vev's of the Higgses and $\mu$ is the supersymmetric Higgs mass.    For simplicity, we fix the values of 
$A_0$ and $\tan\beta$, 
\begin{align}
 A_0 = 0~\mathrm{GeV},\quad \tan\beta = 10,
\end{align}
and take the sign of $\mu$ positive. As we mentioned in Sec.~\ref{sec:introduction}, radiative corrections to the scalar potential is 
minimized at a scale around the extremal value of the up-type Higgs,  $|H_2|_{\mathrm{ext}}$.
Such a case would be an intermediate scale between the weak scale 
and the GUT scale, because 
$|H_2|_{\mathrm{ext}}={\cal O}(\sqrt{m_{3/2}M})$.
Some of our results change significantly around $10^6$ GeV.
Thus, we show numerical calculations for the
RGE scale, $\Lambda$, around  $10^5 - 10^7$ GeV.

\begin{figure}[t]
\begin{tabular}{c}
\includegraphics[width=80mm]{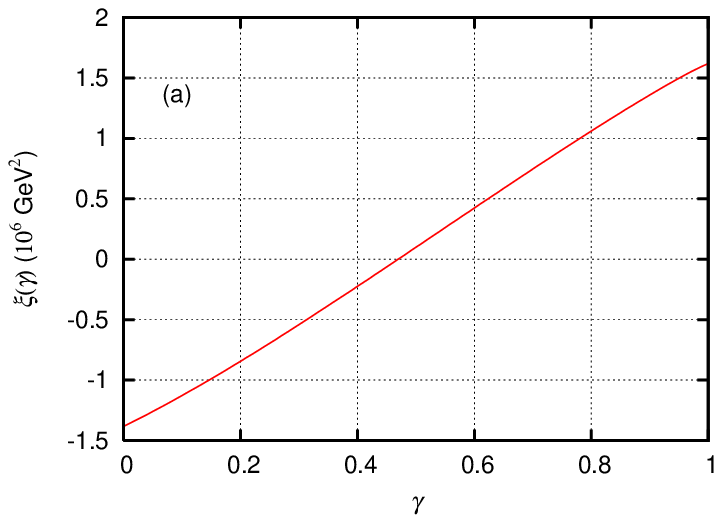} \\ 
\includegraphics[width=80mm]{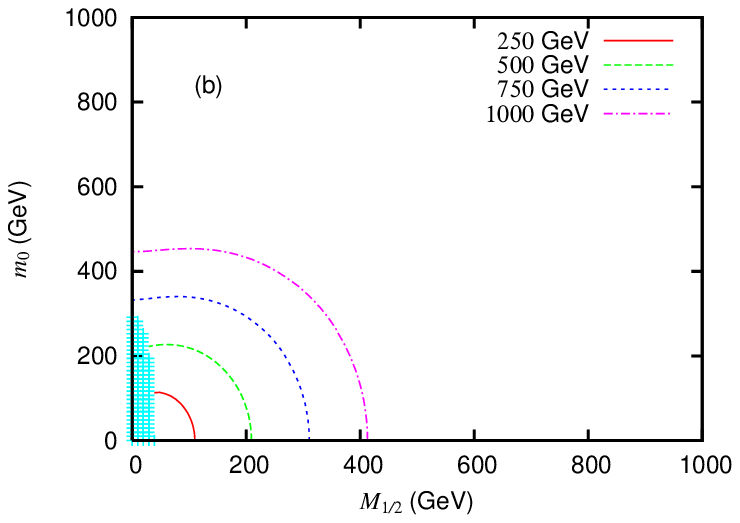} 
\end{tabular}
\caption{Figure (a) shows the function, $\xi(\gamma)$ scaled by $10^{-6}$, and Figure (b) shows the constraint from the MSSM UFB-2. In
 Figure (b), the solid (red), dashed (green), dotted (blue) and the dashed-dotted (pink) curves correspond to $|c'|/|c| = 250,~500,~750$ 
and $1000$ GeV, respectively.}
\label{fig:1}
\end{figure}
In Figure \ref{fig:1}.(a), we plot the function, $\xi(\gamma)$, given in Eq.~(\ref{eq:16}) with $m_0 = 300$ GeV and $M_{1/2} = 500$ GeV at 
$\Lambda = 10^6$ GeV. The function, $\xi(\gamma)$, is an increasing function of $\gamma$, and is always negative at $\gamma = 0$ 
while it is positive at $\gamma=1$ when the EWSB successfully occurs. Hence, $\xi(\gamma)$ has only one zero point for 
$0 \le \gamma \le 1$. The zero point, $\gamma_{\mathrm{ext}}$, is usually found around $0.5$ unless $M_{1/2}$ is small. In the case of 
small $M_{1/2}$, $\gamma_{\mathrm{ext}}$ is found to be $0.4$ or not found because the EWSB does not occur.
The shape of $\xi(\gamma)$ is in general the same for other values of the CMSSM parameters and the RGE scale. 

The constraint from the MSSM UFB-2 direction, Eq.~(\ref{eq:6}), is
shown in Figure \ref{fig:1}.(b) for $\Lambda = 10^6$ GeV. 
We varied $m_0$ and $M_{1/2}$ and 
solved $\xi(\gamma)$ at each point. The solid (red), dashed (green), dotted (blue) and the dashed-dotted (pink) curves represent the constraint with 
$|c'|/|c| = 250,~500,~750$ and $1000$ GeV, respectively. The inside of the curve are excluded by the constraint. It is seen that the
excluded regions expand as $|c'|/|c|$ increases. The hatched (light blue) region is also excluded because the EWSB does not occur due to negative 
$|\mu|^2$.
Results for other RGE scales $\Lambda$ are also the same
qualitatively.

\begin{figure}[t]
\begin{tabular}{c}
\includegraphics[width=80mm]{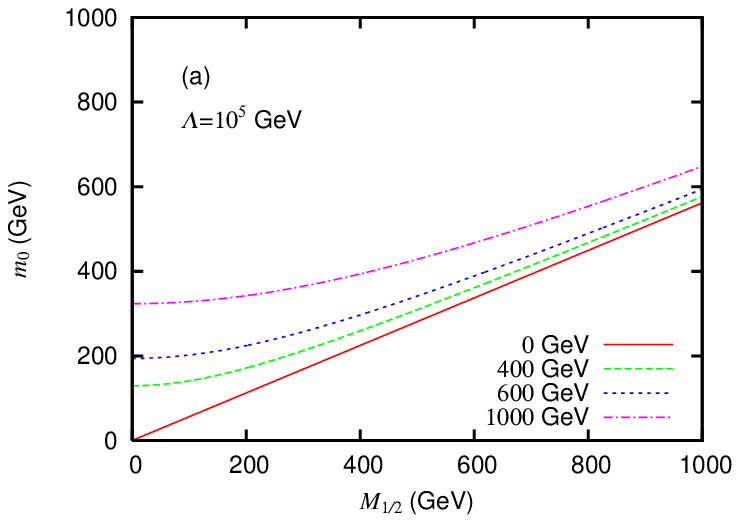} \\ 
\includegraphics[width=80mm]{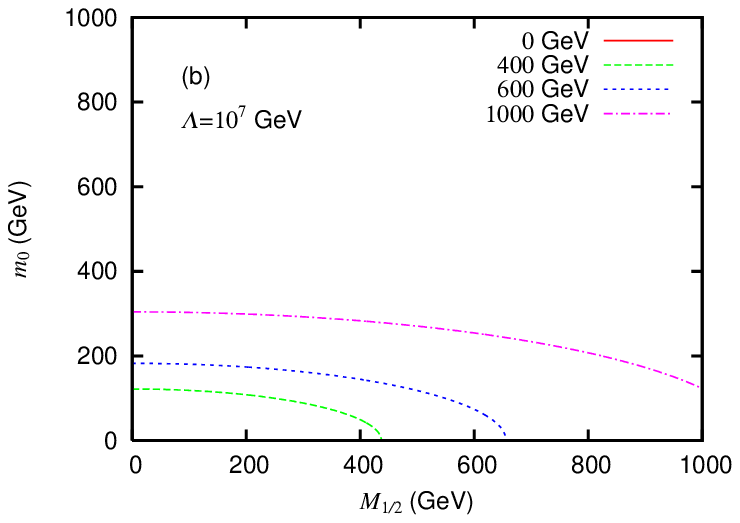} 
\end{tabular}
\caption{The constraint from the MSSM UFB-3 direction is shown in Fig.(a) and Fig.(b) for $\Lambda=10^5$ and $10^7$, respectively. 
The solid (red), dashed (green), dotted (blue) and the dashed-dotted (pink) curves correspond to $|c'|/|c| = 0,~400,~600$ 
and $1000$ GeV, respectively.}
\label{fig:2}
\end{figure}
The constraint from the MSSM UFB-3 direction is shown in \ref{fig:2}.(a) for $\Lambda = 10^5$ GeV and (b) for $\Lambda = 10^7$ GeV. 
The solid (red) curve represents the constraint with the ratio of the coefficients, $|c'|/|c| = 0$ GeV, corresponding to one in the MSSM, 
Eq.~(\ref{eq:8}). The dashed (green), dotted (blue) and dashed-dotted (pink) curves correspond to the ratio of the coefficient, $400,~600$
and $1000$ GeV, respectively. In Fig.2(a), the region below the curves is excluded by the constraint. It is seen that, due to the presence of
the neutrino mass operators, the constraint in the $\nu$SSM is tighter than that in the MSSM. In Fig.2(b), the region inside the curves is 
excluded by the constraint. In the case of $|c'|/|c|=0$ GeV, only the point, $M_{1/2} = m_0 = 0$ GeV, is excluded. It is seen that the 
shape of the excluded region is elliptic in Fig.2(b) while it is
hyperbolic in Fig.2(a). 
The difference originates from the fact that
when we write $m_{H_2}^2(\Lambda)=aM_{1/2}^2+bm_0^2$, 
the coefficient $a$ is negative for $\Lambda \leq {\cal O}(10^6)$ GeV
but the coefficient $a$ is positive for $\Lambda \geq {\cal O}(10^7)$
GeV.
The CCB minimum along the MSSM UFB-3 direction can be negative even if the soft mass of the up-type Higgs is positive since the 
potential is lowered by the term of order $5$th power, which is
proportional to $|c'|$. 
Results for other scales such as $\Lambda \leq {\cal O}(10^6)$ GeV 
and $\Lambda \geq {\cal O}(10^7)$ GeV are the same qualitatively 
as Fig.2(a) and Fig.2(b), respectively.

\section{Summary and Discussion}\label{sec:summary-discussion}
We have considered the $\nu$SSM with neutrino mass operators 
where the  $(\hat{H}_2 \cdot \hat{L})(\hat{H}_2 \cdot \hat{L})$ 
operator in the superpotential and the corresponding dimension 
four operator in the soft SUSY breaking term are added to the MSSM. In this model, the scalar potential contains new terms of order 
$6$th, $5$th and $4$th power which are absent in the minimally supersymmetric extension of the SM. We have analyzed the scalar potential 
along the MSSM UFB and CCB directions and found new unrealistic vacua which appear due to the higher order term in the scalar potential.

We have found that the MSSM UFB directions disappear and turn to be CCB directions due to the presence of higher power terms in the scalar 
potential. The minima along these CCB directions are of ${\cal O}(m_{\mathrm{SUSY}} M)^2$ and can be deeper than that of the EWSB. We have 
derived necessary conditions to avoid the CCB minima along the MSSM UFB-2 and UFB-3 which impose constraints among the soft SUSY breaking 
parameters and the coefficients of the neutrino mass operators. The constraints are expressed in terms of the ratio of the coefficients, 
$|c'|/|c|$, thus these can not be ignored even if $c'$ and $c$ are small. The most stringent constraint was obtained from the MSSM UFB-3 
direction, which imposes bounds on the soft masses of the up-type Higgs and the left-handed sleptons. The constraint holds even if the soft 
mass of the Higgs is positive, therefore it should be always taken into account at any scale.

We have also shown that there appear new CCB minima along the MSSM CCB directions in the case of $|c'|/|Y_u^2| \gg 1$. We showed that 
the extremal value of the Higgs can be determined by neglecting the Yukawa coupling of quarks in this case and the potential can be deeper 
than that of the EWSB. As same as the constraints along the UFB directions, necessary conditions to evade the CCB minima are expressed 
in terms of the ratio of the coefficients. 

In Sec.~\ref{sec:numerical-analysis}, we have applied our results to see differences from the constraints in the MSSM. We calculated the 
soft masses at scales $10^5 \le \Lambda \le 10^7$ GeV using the RGEs in the CMSSM. It was shown that the constraints we derived are more
stringent than those of the MSSM.
Thus, it is important to apply these constraints to 
several SUSY breaking models.
We would study them elsewhere including detailed analysis on 
the CMSSM.

As we have mentioned in the introduction, radiative corrections must be included into analysis for our results to be applied at the 
electroweak scale. It is also needed to include finite temperature effects if one analyze the potential at high energy scale or for large 
vev's. We leave these for our future work.

\acknowledgments
T.~K. is supported in part by the Grant-in-Aid for Scientific Research No.~20540266 and the Grant-in-Aid for the 
Global COE Program "The Next Generation of Physics, Spun from Universality and Emergence" from the Ministry of 
Education, Culture, Sports, Science and Technology of Japan. T.~S is the Yukawa Fellow and the work of T.~S is partially supported 
by Yukawa Memorial Foundation. The work of Yoshimi Kanehata and Yasufumi Konishi is supported by Maskawa Institute at Kyoto Sangyo 
University.

\appendix
\section{Scalar potential}\label{sec:scalar-potential}
In this Appendix, we give notations of scalars and the scalar potential with the neutrino mass 
operators. Throughout this article, flavour indexes are suppressed for simplicity.

The down-type and the up-type Higgs scalars are denoted as 
\begin{align}
 H_1 =
  \begin{pmatrix}
   H^1_1 \\
   H^2_1
  \end{pmatrix},\quad 
 H_2 = 
  \begin{pmatrix}
   H^1_2 \\
   H^2_2
  \end{pmatrix},
\end{align}
where $H_1^1$ and $H_2^2$ are electrically neutral components. Throughout this article, we refer $H^1_1$ and $H^2_2$ as $H_1$ and $H_2$. The left-handed squarks and the right-handed squarks are  
denoted as
\begin{align}
 \tilde{Q} =
  \begin{pmatrix}
   \tilde{u}_L \\
   \tilde{d}_L
  \end{pmatrix},\quad
 \tilde{u}_R,~\tilde{d}_R,
\end{align}
and the left-handed sleptons and the right-handed sleptons are denoted as 
\begin{align}
 \tilde{L} =
  \begin{pmatrix}
   \tilde{\nu}_L \\
   \tilde{e}_L
  \end{pmatrix},\quad
 \tilde{e}_R.
\end{align}

The superpotential is given
\begin{align}
 \mathcal{W} = \mathcal{W}_{\mathrm{MSSM}} 
  -\frac{1}{2} c (\hat{H}_2 \cdot \hat{L}) (\hat{H}_2 \cdot \hat{L}),\label{eq:1}
\end{align}
where $c$ is a coefficient. $\mathcal{W}_{\mathrm{MSSM}}$ is the superpotential of the MSSM,
\begin{align}
 \mathcal{W}_{\mathrm{MSSM}} &= \mu\hat{H}_1\cdot\hat{H}_2 + Y_d \hat{H}_1\cdot \hat{Q} \hat{D}^c_R 
                              + Y_u \hat{H}_2\cdot \hat{Q} \hat{U}^c_R \nonumber \\
                             &\quad + Y_e \hat{H}_1\cdot \hat{L} \hat{E}^c_R,
\end{align}
where $\mu$ is a supersymmetric Higgs mass, and $Y_{u,d,e}$ are the Yukawa couplings of up quarks, down quarks and 
charged leptons, respectively. 

The scalar potential, $V$, is divided into three parts which consist of $F$, $D$ and the soft SUSY breaking terms,
\begin{align}
 V = V_F + V_D + V_{\mathrm{soft}}.
\end{align}
The $F$ term potential, $V_F$, is given by a sum of absolute square of all matter auxiliary fields,
\begin{align}
 V_F = \sum_{i=\mathrm{matter}} |F_i|^2,
\end{align}
where
\begin{subequations}
\begin{align}
 F_{H^1_1}^\ast & = \mu H^2_2 + Y_e \tilde{e}_L \tilde{e}_R^\ast + Y_d \tilde{d}_L \tilde{d}_R^\ast ,\\
 F_{H^2_1}^\ast & = -\mu H^1_2 - Y_e \tilde{\nu}_L \tilde{e}_R^\ast - Y_d \tilde{u}_L \tilde{d}_R^\ast,\\
 F_{H^1_2}^\ast & = -\mu H^2_1 + Y_u \tilde{d}_L \tilde{u}_R^\ast 
    - c \tilde{e}_{L} (H_2^1 \tilde{e}_L - H_2^2 \tilde{\nu}_L), \\
 F_{H^2_2}^\ast & = \mu H^1_1 - Y_u \tilde{u}_L \tilde{u}_R^\ast 
    + c \tilde{\nu}_{L} (H_2^1 \tilde{e}_L - H_2^2 \tilde{\nu}_L),\\
 F_{\tilde{e}_R}   & = Y_e (H^1_1 \tilde{e}_L - H^2_1 \tilde{\nu}_L),\\
 F_{\tilde{e}_L}^\ast   & = Y_e H^1_1 \tilde{e}_R^\ast
    - c H_2^1 (H_2^1 \tilde{e}_L - H_2^2 \tilde{\nu}_L), \\
 F_{\tilde{\nu}_L}^\ast & = -Y_e H^2_1 \tilde{e}_R^\ast
    + c H_2^2 (H_2^1 \tilde{e}_L - H_2^2 \tilde{\nu}_L), \\
 F_{\tilde{d}_R}   & = Y_d (H^1_1 \tilde{d}_L - H^2_1 \tilde{u}_L),\\
 F_{\tilde{d}_L}^\ast   & = Y_d H^1_1 \tilde{d}_R^\ast + Y_u H^1_2 \tilde{u}_R^\ast,\\
 F_{\tilde{u}_R}      & = Y_u (H^1_2 \tilde{d}_L - H^2_2 \tilde{u}_L),\\
 F_{\tilde{u}_L}^\ast & = -Y_d H^2_1 \tilde{d}_R^\ast - Y_u H^2_2 \tilde{u}_R^\ast.
\end{align}
\end{subequations}

The $D$ term potential, $V_D$, is given by a sum of square of all gauge auxiliary fields,
\begin{align}
 V_D = \frac{1}{2} \bigg( (D^a_{SU(3)})^2 + (D^a_{SU(2)})^2 + (D_{U(1)})^2 \bigg),
\end{align}
where $a$ runs from $1$ to $8~(3)$ for $SU(3)~(SU(2))$ and summation over $a$ should be understood. The gauge 
auxiliary fields are given by  
\begin{subequations}
\begin{align}
 D^a_{SU(3)} &= g_3 \left( \tilde{Q}^\dagger \frac{\lambda^a}{2} \tilde{Q} 
                          - \tilde{u}^\ast_R \frac{\lambda^a}{2} \tilde{u}_R
                          - \tilde{d}^\ast_R \frac{\lambda^a}{2} \tilde{d}_R \right), \\
 D^a_{SU(2)} &= g_2 \big( \tilde{Q}^\dagger T^a \tilde{Q} + \tilde{L}^\dagger T^a \tilde{L}
                + H_1^\dagger T^a H_1 + H_2^\dagger T^a H_2 \big),\\
 D_{U(1)} &= g_1 \left( \frac{1}{6} \tilde{Q}^\dagger \tilde{Q} -\frac{2}{3} \tilde{u}_R^\ast \tilde{u}_R 
           + \frac{1}{3} \tilde{d}^\ast_R \tilde{d}_R -\frac{1}{2}\tilde{L}^\dagger \tilde{L}\right. \nonumber \\ 
           &\qquad \left. + \tilde{e}_R^\ast \tilde{e}_R -\frac{1}{2}H_1^\dagger H_1 + \frac{1}{2}H_2^\dagger H_2 \right),
\end{align}
\end{subequations}
where $g_i~(i=1,2,3)$ is a gauge coupling constant, and $\lambda^a$ and $T^a$ are Gell-Mann and Pauli matrix, respectively.

The soft SUSY breaking term, $V_{\mathrm{soft}}$, is given as
\begin{align}
  V_{soft} & = m^2_{H_1} H_1^\dagger H_1 + m^2_{H_2} H_2^\dagger H_2 
             + \big(B\mu H_1 \cdot H_2 + h.c.\big) \nonumber \\
          &~~+ m^2_{\tilde{Q}} \tilde{Q}^\dagger \tilde{Q} + m^2_{\tilde{u}_R} \tilde{u}_R^\ast \tilde{u}_R
             + m^2_{\tilde{d}_R} \tilde{d}_R^\ast \tilde{d}_R \nonumber \\
          &~~+ m^2_{\tilde{L}} \tilde{L}^\dagger \tilde{L} + m^2_{\tilde{e}_R} \tilde{e}_R^\ast \tilde{e}_R  \nonumber \\
          &~~+ \big(A_d Y_d H_1 \cdot \tilde{Q} \tilde{d}_R^\ast + A_u Y_u H_2 \cdot \tilde{Q} \tilde{u}_R^\ast \nonumber \\
          &\qquad + A_e Y_e H_1 \cdot \tilde{L} \tilde{e}_R^\ast + h.c.\big) \nonumber \\
          &~~ - \frac{1}{2} \left( c' (H_2 \cdot \tilde{L}) (H_2 \cdot \tilde{L}) + h.c. \right),
\end{align}
where $m_i~(i=H_1, H_2, \tilde Q, \uR, \dR, \tilde L, \tilde e_R)$ are soft masses and $B\mu$ is a soft term for Higgses. A symbol ``dot'' represents an 
inner product for $SU(2)$ doublets, $A \cdot B = A^1B^2 - A^2 B^1$. The trilinear terms, $A_i~(i=u,d,e)$, are defined to be 
proportional to the corresponding Yukawa coupling. 
We also use the following notations,
\begin{subequations}
\begin{align}
 m_1^2 &= m_{H_1}^2 + |\mu|^2,\\
 m_2^2 &= m_{H_2}^2 + |\mu|^2,\\
 m_3^2 &= - B \mu.
\end{align}
\end{subequations}

\section{General form of the vacuum expectation values}\label{sec:general-form-vacuum}
We give the general form of the vacuum expectation value of $H_2$ and show that the extremal value of $|H_2|$ is independent of $Y_u$ at the
leading order when we assume that $|c'|/|Y_u|^2 \gg 1$.

Firstly, we give the general form of the vev of $|H_2|$. Differentiating the potential, (\ref{eq:27}) with respect to $|H_2|$, the equation 
to be solved is obtained,
\begin{align}\label{eq:B1}
6 \hat{C} |H_2|^4 -  5 \hat{D} |H_2|^3 + 4 \hat{F} |H_2|^2 - 3 \hat{A} |H_2| + 2 \hat{m}^2 = 0,
\end{align}
where the dependence of coefficients 
on $\alpha,~\beta,~\gamma$ and $\gamma_L$ are omitted.
We introduce a dimensionless parameter $x$ which is defined by
\begin{align}
 x \equiv \frac{|H_2|}{M},
\end{align}
where $M$ is the cut-off scale of the neutrino mass operators.
Then, Eq.~(\ref{eq:B1}) is written as
\begin{align}
 a_4 x^4 + a_3 x^3 + a_2 x^2 + a_1 x + a_0 = 0,\label{eq:28}
\end{align}
where
\begin{subequations}
\begin{align}
 a_4 &= 6 \hat{C} M^2, \\
 a_3 &= -5 \hat{D} M, \\
 a_2 &= 4 \hat{F}, \\
 a_1 &= -3 \hat{A} M^{-1}, \\
 a_0 &= 2 \hat{m}^2 M^{-2}.
\end{align}
\end{subequations}
The equation, (\ref{eq:28}), can be deformed to the following form by shifting $x = t + t_0$ with $t_0 = - a_3/(4a_4)$,
\begin{align}
 t^4 + b_2 t^2 + b_1 t + b_0 =0,\label{eq:29}
\end{align}
where
\begin{subequations}
\begin{align}
 b_2 &= 6 t_0^2 + 3 a_3 t_0/a_4 + a_2/a_4, \\
 b_1 &= 4 t_0^3 + 3 a_3 t_0^2/a_4 + 2 a_2 t_0/a_4 + a_1/a_4, \\
 b_0 &= t_0^4 + a_3 t_0^3/a_4 + a_2 t_0^2/a_4 + a_1 t_0/a_4 + a_0/a_4.
\end{align}
\end{subequations}
We rewrite Eq.~(\ref{eq:29}) as
\begin{align}
 \left( t^2 + \frac{b_2}{2} + u \right)^2 &= 2 u \left( t - \frac{b_1}{4u} \right)^2 \nonumber \\ 
  &\quad - \frac{b_1^2}{8 u} + \left( \frac{b_2}{2} + u \right)^2 - b_0.
\end{align}
The above equation has the solution 
\begin{align}
 t = \frac{1}{2} \left( \mp \sqrt{2u} \pm \sqrt{2u - 2 ( b_2 + 2 u \mp \frac{b_1}{\sqrt{2u}})} \right),
\end{align}
if $u$ satisfies the following equation,
\begin{align}
 - \frac{b_1^2}{8 u} + \left( \frac{b_2}{2} + u \right)^2 - b_0 = 0. \label{eq:30}
\end{align}
The equation, (\ref{eq:30}), is also rewritten as 
\begin{align}
 u^3 + c_2 u^2 + c_1 u + c_0 = 0,
\end{align}
where 
\begin{subequations}
\begin{align}
 c_2 &= b_2, \\
 c_1 &= \frac{b_2^2}{4} - b_0, \\
 c_0 &= - \frac{1}{8}b_1^2,
\end{align}
\end{subequations}
and it is deformed by shifting $u = s + s_0$ with $s_0 = - c_2/3$,
\begin{align}
 s^3 + d_1 s + d_0 = 0, \label{eq:31}
\end{align}
where
\begin{subequations}
\begin{align}
 d_1 &= 3 s_0^2 + 2 c_2 s_0 + c_1, \\
 d_0 &= s_0^3 + c_2 s_0^2 + c_1 s_0 + c_0.
\end{align}
\end{subequations}
The solutions of the equation, (\ref{eq:31}), are well-known and given by 
\begin{align}
 s = p + q,
\end{align}
where
\begin{align}
 (p,q) = (p_0,q_0),~(p_0 \omega, q_0 \omega^2),~(p_0 \omega^2, q_0 \omega),
\end{align}
and
\begin{subequations}
\begin{align}
 p_0 &= \sqrt[3]{-\frac{d_0}{2} + \sqrt{ \frac{d_0^2}{4} + \frac{d_1^3}{27} } }, \\
 q_0 &= \sqrt[3]{-\frac{d_0}{2} - \sqrt{ \frac{d_0^2}{4} + \frac{d_1^3}{27} } }, \\
 \omega &= \frac{1}{2}( -1 + \sqrt{3}i).
\end{align}
\end{subequations}

Now we are in a position to show that the leading order of $|H_2|_{\mathrm{ext}}$ is independent of $Y_u$. We assume that 
\begin{align}
 &\frac{|c'|}{|Y_u|^2} \gg 1, \\
 &|c'| \simeq \frac{m_{\mathrm{SUSY}}}{M},
\end{align}
where $m_{\mathrm{SUSY}}$ is the SUSY breaking scale and $|Y_u|$ is of order $10^{-5}$ which corresponds to the up quark mass. 
Then, the coefficients, $a_i,~(i=0-3)$, are of order 
\begin{align}
 a_3 & \sim  |Y_u|,\\
 a_2 & \sim \frac{m_{\mathrm{SUSY}}}{M}, \\
 a_1 & \sim |Y_u| \frac{m_{\mathrm{SUSY}}}{M}, \\
 a_0 & \sim \left( \frac{m_{\mathrm{SUSY}}}{M} \right)^2.
\end{align}
Using this order estimation, we can estimate the $u$ and $t$,
\begin{align}
 u \sim t \sim \frac{m_{\mathrm{SUSY}}}{M},
\end{align}
since $b_1$ and $b_2$ are of order,
\begin{align}
 b_1 &\sim |Y_u| \frac{m_{\mathrm{SUSY}}}{M}, \\
 b_2 &\sim \left( \frac{m_{\mathrm{SUSY}}}{M} \right)^2.
\end{align}
Thus, $x$ can be estimated as
\begin{align}
 x \sim \sqrt{ \frac{m_{\mathrm{SUSY}}}{M} }.
\end{align}
The leading order of $|H_2|_{\mathrm{ext}}$ is determined by $m_{\mathrm{SUSY}}$ and $M$, and independent of $|Y_u|$.  
This result implies that we can neglect $\hat{D}$, $\hat{A}$ and the term proportional to $|Y_u|$ in $\hat{F}$ to obtain $|H_2|_{\mathrm{ext}}$.

\bibliographystyle{apsrev}
\bibliography{biblio}

\end{document}